# 2D Au@PNiPAM microgel arrays suitable for photonic devices with thermally controlled interparticle gap

Joaquim Clara Rahola, Rafael Contreras Cáceres, Antonio Fernandez Barbero

*Dated: September 27$^{th}$ 2012*

Core-shell Au@pNIPAM nanocomposites are assembled through a simple and inexpensive approach where drops of bulk nanoparticle solution are left to dry on ITO surfaces at fixed and finely tuned temperatures. In this process particles deposit and exhibit predominant hexagonal structure with interparticle distances determined by the length of the thermoresponsive PNiPAM shell. This process allows controlling the distance among gold cores as function of the drying temperature with the PNiPAM shell acting as mechanical spacer. Structural characterization is performed through microscope imaging techniques. We locate the coordinates of each deposited particle and quantify the pair and orientational correlation functions characteristic to each array at each deposition temperature. The elasticity of the polymer shell at each temperature is considered as the shell is more elastic at temperatures below PNIPAM's LSCT while the particle resembles more a hard-sphere colloid at higher temperatures. These two factors, thermal PNiPAM response and shell elasticity, entirely determine the ensemble structure as well as the interparticle gap. In a second step, previously prepared Au@PNiPAM arrays are exposed to atmospheric plasma. Plasma reacts and degrades the PNiPAM shell, being non-invasive to Au cores which remain at their original array positions. 2D ordered arrays of bare gold nanospheres are achieved with the center-to-center distance previously tuned by temperature which lays promising science and technological applications.

## 1. Introduction

Metal nanoparticles exhibit optical and spectroscopic properties which differ remarkably from those from bulk materials, one of them being the well-known Localized Surface Plasmon Resonance (LSPR)[1]. This phenomenon is produced when an external electromagnetic field incident on a metal nanoparticle induces a delocalization of the electron cloud. A net charge difference on the nanoparticle surface acts as restoring force and in the simplest case produces a dipolar oscillation. Optical responses are originated from strong metal nanoparticle absorption when the frequency of the electromagnetic field becomes resonant with the coherent electron motion[2]. Coupled metal nanoparticles at close distances display intense optical fields concentrated at the interparticle gap, and broad spectral tenability is achieved by simply varying this gap[3]. A wide variety of applications are based on the fabrication of regular surface patterns with controlled interparticle distances at the nano length scale, becoming this fact crucial for applications such as molecular sensing, optical filtering or subwavelength lithography[4]. The local regularity of those structures is of paramount importance for collective plasmonic properties which are mainly determined by short-range

*Group of Complex Fluids Physics, University of Almeria, 04120, Spain.*

[quimtxo@ual.es](mailto:quimtxo@ual.es), [rcc689@ual.es](mailto:rcc689@ual.es), [afernand@ual.es](mailto:afernand@ual.es)

couplings. Among a vast set of plasmonic nanosystems, self-assemble of metal nanoparticles to form periodic arrays provides a cheap and effective bottom-up approach for the mass production of large-area plasmonic devices[5, 6]. In the last decades, 2D particle arrays have been successfully obtained using different strategies as lithography based, solvent evaporation, miniemulsions loaded with a metal precursor complex, spin-coating or Langmuir-Blodgett methods where the distance between the particles is controlled by electrostatic and interfacial forces, dramatically sensitive to the deposition conditions and to the presence of contaminants[2, 7-12]. A simple alternative to those methods is to use tunable mechanical spacers located in between the particles. This is a reliable strategy with a significant reduction of the control parameters. In this context, smart polymers are excellent candidates, being the particle gap set by the polymer swelling state. Recently hybrid Au@PNiPAM nanoparticles have been employed to generate gold surface patterns[8, 13, 14]. Here Au@PNiPAM core-shell nanoparticles are synthesized with different shell-sizes and assembled on silicon wafers, being necessary a new particle synthesis for each preselected interparticle distance and consequently, gap tuning turns necessarily discrete and poorly precise.

In this paper we take advantage of the intrinsic temperature response of the PNiPAM shell and present an inexpensive and relatively easy method for creating 2D arrays of nanoparticles with a well defined interparticle distance. An initial suspension of Au@PNiPAM nanoparticles is carefully dried at fix temperature, which is also carefully controlled. The thermoresponsive nature of the PNiPAM shell allows interparticle distance matching in a continuous fashion with high precision. Using this new method, it is necessary to synthesize only one batch of nanoparticles in order to create structures with different characteristic gaps, being the deposition temperature the only control parameter. Structural features were studied in depth once the arrays formed, with special attention to the role of temperature on the microgel spatial and orientational configuration. We synthesize and deposit Au@pNIPAM hybrid nanoparticles with a 64 nm spherical Au cores covered by a PNiPAM hydrogel shell. The metal core provides an optical plasmon resonance, while the PNiPAM shell acts as mechanical spacer. PNIPAM shows a Lower Critical Solution Temperature (LCST) at *32ºC*. In aqueous suspension and at temperatures way below the LCST, PNiPAM bonds with hydrogen are dominant against PNiPAM-PNiPAM interactions, the shell is hydrophilic and swells. However, when the temperature is raised, hydrogen bonds gradually break down, PNiPAM-PNiPAM interactions become preferential and the shell shrinks. Such collapse reaches down a minimal length once the critical temperature is crossed as PNiPAM chains display a transition from a coil to a globular configuration at the LCST[15, 16]. The temperature is the parameter determining the length of the PNiPAM brush, which in turn controls the gap between Au cores. Imaging techniques reveal that the deposited particle arrays exhibit spatial and angular structure with variable interparticle gap upon temperature settings. Finally, initial particle arrays are modified through controlled removal of the PNiPAM shells by plasma etching at non-invasive conditions to the Au cores. Thus we present an easy and inexpensive protocol for generating nanoparticle arrays potentially useful in multiple science and technological applications such as Surface Enhanced Raman Spectroscopy (SERS) or nanosensor devices.

## 2. EXPERIMENTAL
### 2.1 Materials:
Cetyltrimethylammonium bromide (CTAB), 3-butenoic acid, N-Isopropylacrylamide (NIPAM, 97%) and Indium Tin Oxide (ITO) coated glass slide with 30-60 Ω/sq surface resistivity were purchased to Aldrich. Tetrachloroauric acid ($HAuCl_4 x 3H_2O$), trisodium citrate dihydrate were supplied by Sigma. N.N´-Methylenebisacrylamide was supplied by Fluka. 2,2´-Azobis(2-methylpropionamidine) dihydrochloride was supplied by Acros Organics. All reactants were used without further purification. Milli-Q water with resistivity higher than 18.2 MΩ cm was used in all experiments.

### 2.2 Experimental set-ups
Scanning Electron Microscopy (SEM) images are obtained with a Hitachi microscope model S-3500N operating at 10.0 kV for secondary-electron imaging. The bulk particle size and volume transition of the Au@pNIPAM colloidal particles along the swelling process was monitored by Dynamic Light Scattering (DLS) employing a Malvern 4700 setup with a 5 mW helium-neon laser, λ=632.8 nm, operating at 60º scattering angle. The temperature was controlled to a precision of ±0.1ºC using a Peltier temperature control system aided by external water flow. The mean diffusion coefficient was derived from the intensity autocorrelation function using cumulant analysis and converted into mean particle size though the Stokes-Einstein relation[17]. AFM studies were performed using a Veeco Innova Scanning Probe Microscope operating in tapping-mode with a 0.9 $Nm^{-1}$ spring constant cantilever. We also employ an Atomflo 400D Atmospheric Plasma System from Surfx Technologies. The instrument is equipped with an Atomflo 400 Controller, an x-y-z robot and an AH-250D plasma torch applicator.

### 2.3 Synthesis and characterization of Au@pNIPAM nanoparticles:
Spherical Au nanoparticles with 64 nm diameter are used as seed particles for a subsequent encapsulation with N-Isopropylacrylamide. These gold particles are obtained by variation of the seed mediated method. 35 mL of approximately 15 nm Au nanoparticles 0.5 mM (previously obtained by conventional citrate reduction), are mixed with 15 mL of 0.03 M CTAB aqueous solution. 4.5 mL of this colloidal solution are added while gentle magnetic stirring to 50 mL of a growth solution containing 1mM $HAuCl_4$, 0.015 M CTAB and 800 µL of 3-butenoic acid at 70ºC. After 10 minutes, the solution is centrifuged at 4500 rpm for 40 min to remove the excess of 3-butenoic acid. The pellet is then redispersed in 50 mL of 4 mM CTAB aqueous solution and this dispersion is again centrifugated at 4500 rpm during 30 min to remove the CTAB excess. The final precipitate is redispersed in 10 mL of water. This solution is heated at 70ºC in a 15 mL vial under a $N_2$ flow and 169.8mg of N-Isopropylacrylamide and 23.4mg of N,N´-methylenebisacrylamide are added with gentle magnetic stirring of the system. The subsequent polymerization is initiated 15 min later with 100 µL of 0.1M 2,2´-Azobis(-methylpropionamidine) dihydrochloride. After 2h at 70ºC, the colloidal dispersion is allowed to cool down at room temperature. Finally, in order to remove small oligomers, non-reacted monomers, as well as gold free microgels, the dispersion is diluted to 50 mL, centrifuged at 4000 rpm for 30 min, and redispersed in the same amount of water (fivefold)[13]. DLS measurements reveal the typical temperature transition of the N-Isopropylacrylamide microgel in aqueous solution at *T ~ 32 ºC*, with particle hydrodynamic diameter at bulk conditions ranged between 217nm, at the swollen state, and 355 nm at the collapsed one. Thus particles exhibit a swelling ratio of 0.61.

**2.4 Deposition process**

ITO surfaces are cut to a final size of approximately 1cm$^2$ and cleaned by immersing them in acetone while sonicated for 20 minutes. Then they are transferred to a water solution and sonicated for another 15 minutes. Finally, the ITO substrates are dried with a N$_2$ flow. 20 µL of the Au@pNIPAM colloidal solution with 0.008wt% particle concentration are dropped onto the previously cleaned ITO surfaces, the system is dried in a preheated furnace (Binder, model ED53 RS422) at fixed temperature. We prepared samples at *25ºC, 28ºC, 30ºC, 32ºC, 34ºC, 36ºC* and *38ºC*. The samples were left drying in the furnace for 2 hours.

**3. RESULTS AND DISCUSSION**

**3.1 Assembly of Au@PNiPAM nanoparticle arrays**

Hybryd Au@PANiPAM particles are deposited on ITO flat surfaces and let dry at a fixed and well controlled temperature as described in the previous section. At different deposition temperatures, the response of the PNiPAM shell determines its thickness. This length is the parameter responsible for setting the center-to-center interparticle gap in all deposition processes. Through this strategy, we aim to assemble Au@PNiPAM nanoparticles into 2D arrays as indicated in Fig. 1a. Such protocol is designed so that at low temperatures a large gap is observed in contrast with the short one at temperatures above the LCST. Note that in between these two limiting separation lengths, the interparticle gap can be of any length provided fine temperature tuning as polymer swelling proceeds continuously. AFM imaging reveals such particle deposition at *25ºC* which exhibits larger interparticle distance when compared with AFM images of Au@PNiPAM particles deposited at *36ºC* as observed in Fig. 1b and Fig. 1c. The core-shell morphology is visualized as two distinct bright levels are resolved, one corresponding to the Au cores at the centre and another one corresponding to the polymer shells. AFM imaging also indicates that particles structure into a predominant hexagonal conformation, with the interparticle distance determined by the thickness of the shell. Fig. 1b, Fig. 1c and their insets evidence shell deformations from spherical to hexagonal shapes due to particle overcrowd.

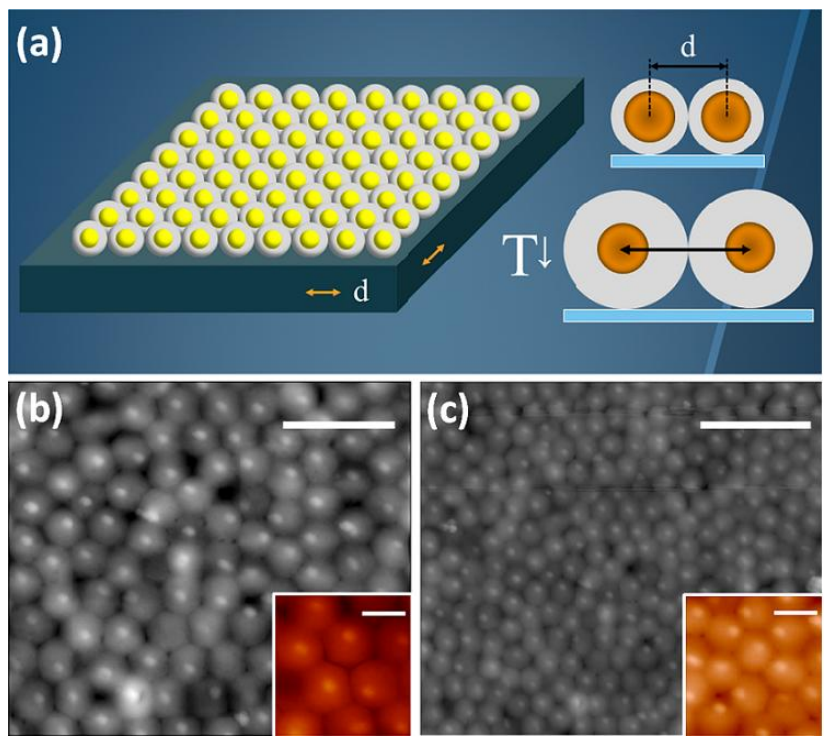

Fig. 1. Schematic description of the 2D Au@PNiPAM arrays after the deposition protocol (a) and AFM images of particle films deposited at *25ºC* (b) and at *36ºC* (c). The thickness of the temperature controlled PNiPAM shell determines the interparticle gap and most particles exhibit nearest neighbors in hexagonal configuration. The scale bars are 1μm in length. The insets in (b) and (c) are magnified AFM images where shell deformation due to particle overcrowd is visualized. Note the brighter Au cores at the center. The inset scale bars are 300nm.

**3.2 Spatial configuration of Au@PNiPAM nanoparticle arrays**

To better understand the structural properties of the assembled hybrid Au@pNIPAM particles at the preset deposition temperatures, we image our arrays employing the SEM set-up described in the experimental section. Due to the high conductivity of Au when compared to PNiPAM, the images display stronger contrast for the Au cores than the PNiPAM shells as observed in Fig. 2(a) and Fig. 2(b), which respectively belong to particles deposited at *25ºC* and *36ºC*. Moreover, in the field of view achieved through SEM, predominant hexagonal ordering is displayed as well as a closest interparticle distance at high than at low temperatures. We study the spatial configuration of deposited particles through particle identification methods typically used in colloidal particle tracking[18]. We developed in-house software in order to filter the gray scaled SEM images and through discriminating false particles by their brightness and/or odd radii, we end up identifying the position of each Au core as indicated in Fig. 2(a) and Fig. 2(b). The radial particle pair correlation function *g(r)* is calculated with *r* being the centre-to-centre distance between normalize the particle correlation function, we set the magnitude of *g(r)* at the largest *r* to be *g(r >>) = 1*, as at such distance the probability of finding a particle is constant. In order to achieve significant statistics, we take 15 SEM images at each temperature and find the corresponding *g(r)*. The final radial particle correlation function is the average from the 15 resulting radial correlation functions from each image. Typical radial correlation functions are displayed in Fig. 3 where all *g(r)* exhibit a correlation primary peak. As *g(r)* quantifies the probability of finding a particle at distance *r* from a test one, such peak indicates the preferential positioning of particles at a particular distance $r_{peak}$ around other ones. The secondary peaks that also characterize radial correlation functions, as seen in Fig. 3, are located at distances multiple of $r_{peak}$ and therefore these hallmarks indicate that particles deposit into a structured configuration.

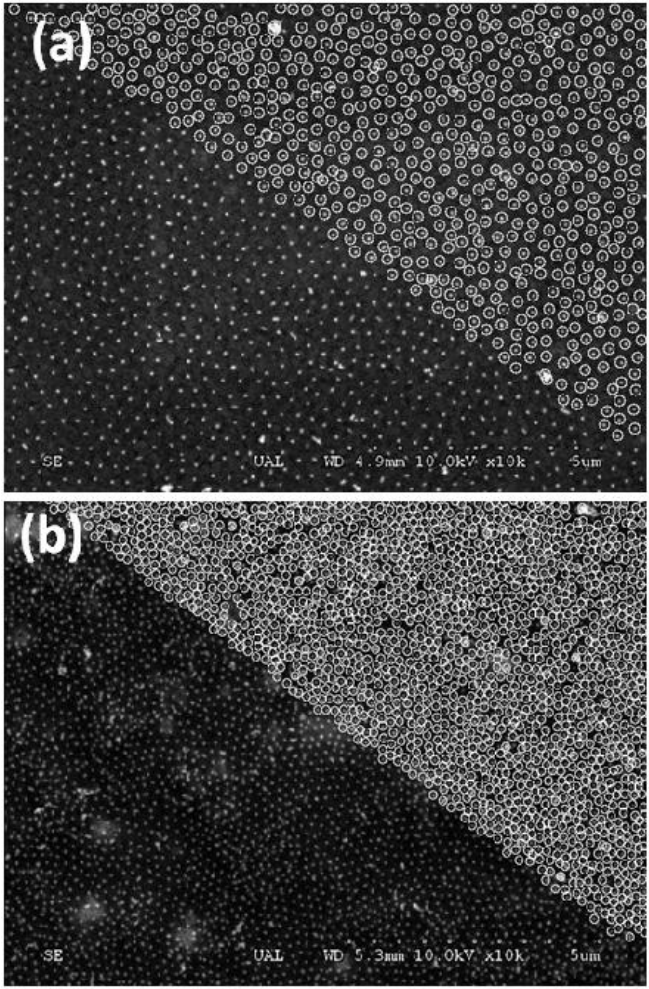

Fig. 2. SEM images of deposited Au@PNiPAM particle arrays at *25ºC* (a) and *36ºC* (b). Bright spots belong to Au cores and circles indicate identified particles.

particles. Concentric rings with thickness $\delta$ and inner radius $r-\delta/2$ are generated and the number of neighboring particles located in each ring is considered. The particle density is the number of particles $\rho$ over the area of one ring, $\rho/[N\pi[(r+\delta/2)^2-(r-\delta/2)^2]] = \rho/(2\pi rN\delta)$, normalized over the number of detected particles *N*. Then the radial particle correlation function is calculated as $g(r) = \langle \rho/[2\pi r N\delta]\rangle$, where the brackets denote the average over all detected particles. To properly

The peak positions of each *g(r)* scale onto a unique master curve by a mere linear shift of the distance *r* to a scaled distance $r_{scaled}$, i.e. $r_{scaled} = \alpha r$, displayed in Fig. 4(a). However, the scaling constant $\alpha$ is not linearly related to the hydrodynamic diameter $R_h$, as shown in Fig. 4(b). In contrast, $\alpha$ and $R_h$ are related through two linear regions separated by the LCST. If interparticle gaps were entirely determined by the length of the PNiPAM shell in bulk, we would expect a linear dependence between $\alpha$ and $R_h$ as the deposition mechanism is the same through all temperatures and deswelling of the shell is continuous. Seemingly, when comparing the response of the bulk DLS particle diameter, $d_h$, with temperature to the one of the primary peak position of the pair correlation functions, $d_{peak}$, it is remarkably observed how $d_{peak}$ responds to temperature in the same fashion than $d_h$ by displaying a transition to a minimal particle size at the same LCST as determined by DLS experiments, shown in Fig. 4(c). However, the magnitude of the closest interparticle distance is lower than the bulk hydrodynamic size at low temperatures and these two lengths get gradually closer as the temperature is raised to finally converge at $T \geq LCST$.

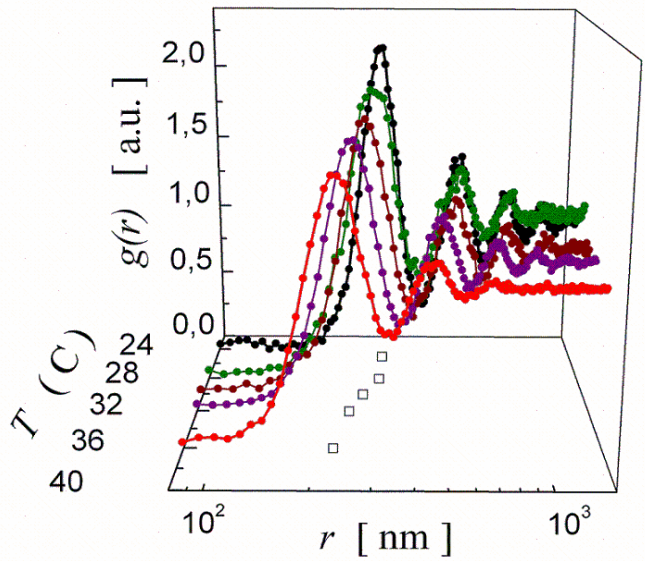

Fig. 3. Radial pair correlation functions $g(r)$. Red symbols belong to *36ºC*, purple symbols to *32ºC*, brown symbols to *30ºC*, green symbols to *28ºC* and black symbols belong to *25ºC*. Projection of the primary peak on the temperature-interparticle distance plane, indicated by black open squares, hints the gap response to temperature.

These features indicate that the PNiPAM response to temperature is the parameter tuning the center-to-center particle positions, however, as such lengths are systematically lower than the bulk DLS particle diameter at T < LCST, it is tempting to suggest that shell elasticity plays an active role in the array formation process. This is the case of multiple microgel systems based on the synthesis of PNiPAM which exhibit new features such as novel phase behaviors or intrasturctural configurations upon presetting the particle elasticity[19-21]. The proposed scenario is the one where particle shells are more elastic at temperatures below the LCST than above this temperature and where particles find a reduced effective surface deposition area at lower temperatures. At low temperatures bigger particles find more difficult to arrange on the same surface area than smaller ones at high temperatures due to a larger particle number concentration, which, in turn, increases the osmotic pressure of the system in the deposition process. Below the LCST the shell is more elastic as it is larger, less dense than at high temperatures and in bulk and it also encapsulates more solvent volume. Then Au@PNiPAM particles deposited at temperatures below the LCST must deform from their bulk size due to particle overcrowd. Such deformation accounts for the difference between the DLS hydrodynamic diameter and the primary peak position of $g(r)$. At temperatures above the LCST, particles are more compact and deformation, despite being present, is not significant. It must be remarked that AFM imaging evidences particles deforming to a hexagonal shape as observed in Fig 1(b) and Fig. 1(c). At *25ºC*, $d_{peak} = 0.947 d_h$, slightly larger than the theoretical prediction of $0.866 d_h$ for completely hexagonal particles. This difference is due to the fact that not all particles are equally deformed since the structure is not perfectly isotropic at large ranges, together with groups of particles only exhibiting partial deformation. As the LCST is approached, deformations are present but less significant. The minimal shell thickness is achieved above the LCST and particle deformation is not dramatically varying the interparticle distance when compared to the closest center-to-center distance in bulk.

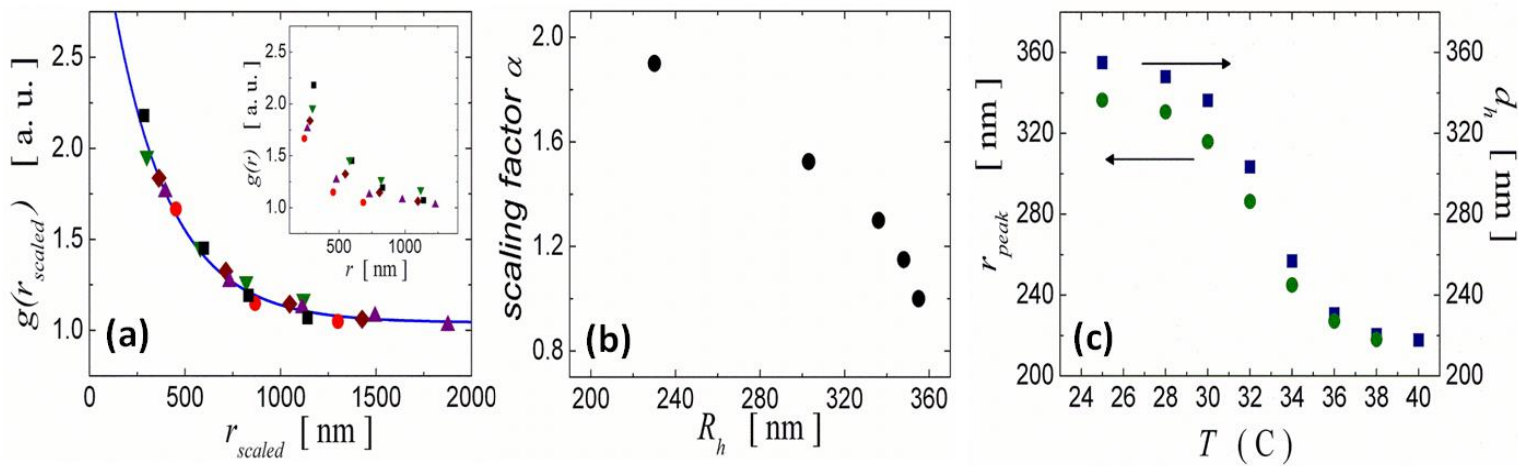

Fig. 4. The peak position of pair correlation functions at different temperatures is scaled by shifting the peak position, $r_{scaled} = \alpha\, r_{peak}$, on top of the peaks belonging to *25ºC* (a). Circles denote *36ºC*, upward triangles *32ºC*, diamonds *30ºC*, downward triangles *28ºC* and squares *25ºC*. The solid line is the fit to an exponential functional form. The inset displays the unscaled data. The scaling factor $\alpha$ gradually decreases with the DLS hydrodynamic diameter revealing two regions between temperatures above and below the LCST (b). The primary peak position versus temperature indicates that the interparticle distance responds to temperature in a close fashion than the DLS hydrodynamic diameter (c). Circles denote the primary peak position while squares indicate the hydrodynamic diameter. Arrows are a visual aid indicating to which axis belongs each dataset.

The scaled master curve, Fig. 4(a), can be well described through an exponential decay $g(r_{scaled}) \sim A\exp(-r/\xi_0)+C$ with the characteristic decay length $\xi_0$ = *302nm*, the intercept *A = 2,67* and *C =1*. By reverse scaling, the characteristic length at each temperature, $\xi_0(T)$, is found. We adopt this protocol in order to obtain reliable statistics in the fit. In order to properly compare the different correlation lengths, they are normalized with the peak of the pair correlation function at each temperature $d_{peak}(T)$. As shown in Fig. 5, the temperature evolution of $\xi(T)/d_{peak}(T)$ does not display significant variation. Within the studied temperature range $\xi(T)/d_{peak}(T) \sim 1$. This feature indicates that the only relevant array length scale is the particle diameter and additionally, it points out that array formation is independent of the deposition rate, which is set by the temperature of each drying process.

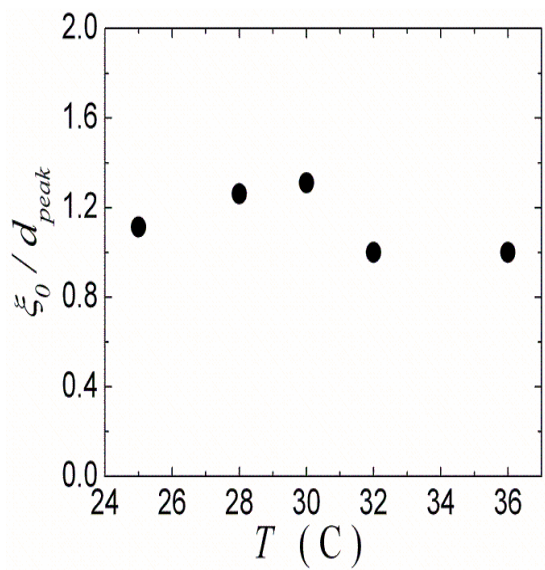

Fig. 5. Normalized characteristic correlation lengths versus temperature.

### 3.3 Orientational configuration of Au@PNiPAM nanoparticle arrays

From particle coordinates, the relative angle between two particles, $\theta_{pq}$, is calculated where $p$ and $q$ denote each particle. Then we find the angle distribution function of neighboring particles with the neighbor condition belonging to those particles located at a distance $r \leq d_{min}$ from the particle taken as reference, with $d_{min}$ the first minimum after $d_{peak}$ in the pair correlation function $g(r)$. This operation is extended to all detected particles at a fixed temperature and the relative angle histogram is found. Then the angle distribution function, $P(\theta)$, is determined by averaging through the number of neighbors and normalizing so that $\int P(\theta) d\theta = 1$. Interestingly for all temperatures the angle distribution function exhibits a peak at $\theta = 0$ and at $\theta = \pi/3$ as shown in Fig. 6(a) and Fig. 6(b) for 25ºC and 36ºC. This feature quantifies arrays displaying preferential hexagonal structure which was also hinted by visual inspection of AFM and SEM images. We then consider the orientation order parameter $\Psi_{6p} = \sum^q e^{6i\theta_{pq}} / N$ and calculate the orientation correlation function $g_6 = \langle \Psi^*_6(r_i) \Psi_6(r_j) \rangle$[22] to all 15

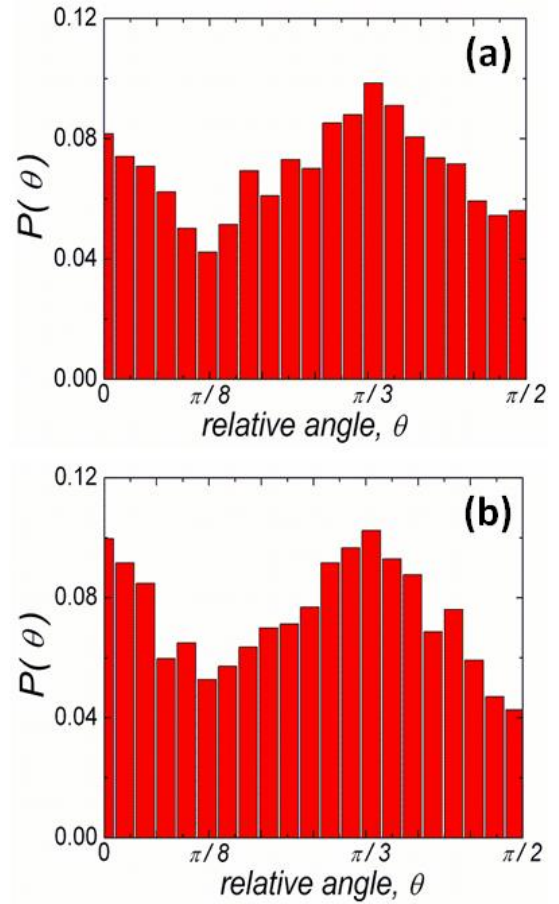

Fig. 6. Relative angle probability distribution functions between neighboring particles at 25ºC (a) and 36ºC (b).

SEM image datasets taken at each temperature. The gradual decrease of $g_6$ indicates how angular order blurs with distance. Such decrease can be described by an exponential functional form so that $g_6 \sim exp\ -r/\xi_6$ with $\xi_6$ the characteristic decay length, indicating the short-ranged angular order over the whole structure. In order to compare the angular correlation functions we normalize all distances to the corresponding interparticle peak $d_{peak}$. Interestingly, the decay length of $g_6$ increases with increasing temperature as shown in Fig 7(a) and the normalized correlation length $\xi_6/d_{peak}$ gradually increases with temperature as indicated in Fig 7(b). $\xi_6/d_{peak}$ increases by a factor of 4 in the studied temperature range, about 9 microgel diameters, and therefore high temperature deposition not only reduces the interparticle gap but also enhances the local orientational order. This is an interesting result since temperature usually promotes disorder. Note however, that by increasing the temperature, shell elasticity decreases, and particles resemble hard-spheres. This suggests that less elastic shells better accommodate particles around others while the deposition process takes place.

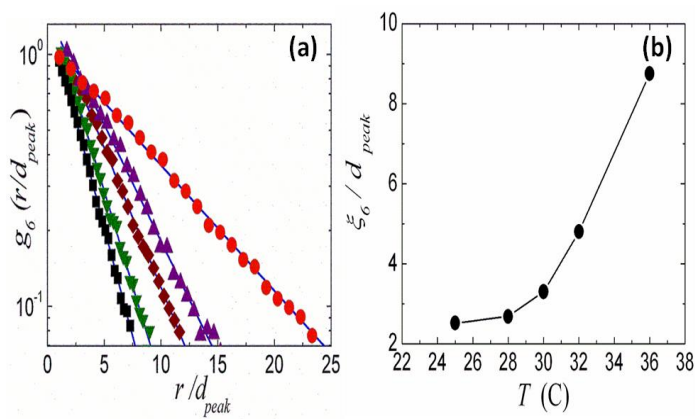

Fig. 7. Angular correlation functions $g_6(r/d_{peak})$ at different deposition temperatures (a). Circles belong to *36ºC*, upward triangles to *32ºC*, diamonds to *30ºC*, downward triangles to *28ºC* and squares to *25ºC*. Note that distances have been normalized by the pair correlation peak of each corresponding temperature, $d_{peak}$. The normalized angular correlation length, $\xi_6/d_{peak}$, versus temperature is displayed in (b).

**3.4 Gold particle arrays with controlled gap**
Once Au@PNiPAM arrays have been successfully achieved, we aim to remove the PNiPAM shell around the cores so that we obtain ordered arrays of Au nanoparticles. Heating procedures are avoided in order to prevent possible swelling or deswelling from the PNiPAM shell and also, in order to respect as much as possible the mechanical properties of the substrate. A new strategy is chosen, where initial arrays are treated by employing atmospheric plasma. Arrays of single gold particles were obtained by removing the polymer used to set the interparticle gap. Oxidation by $Ar/O_2$ 100:1 atmospheric plasma etching was employed at low constant power of 30W. This is a low temperature, fast and cheap procedure, reliable at industrial scale[23]. Fig. 8(a), Fig. 8(b), Fig. 8(c) and Fig. 8(d) show SEM images of treated Au@pNIPAM arrays which belong to a deposition temperature of *25°C* and to different exposure times. Note that the exposure time is the time in which an original array is left under the plasma cloud, i.e. in all plasma exposition experiments we employ a fresh deposit. The spacing between the deposit and the plasma head is 1cm. The control unexposed system is shown in Fig. 8(a). As it can be observed in Fig. 8(b), there is no appreciable polymer oxidation at 20s of exposure. However, partial microgel degradation was obtained at 40s of plasma treatment as displayed in Fig. 8(c). A complete removal of the PNiPAM shell was achieved at 90s of plasma exposure, and as it can be observed in Fig. 8(d). This exposure time did not affect the array structure of the remaining Au nanoparticles. Therefore the hybrid Au@PNiPAM particle deposits were transformed into gold nanoparticle arrays by controlled removal of the microgel shell employing the atmospheric plasma treatment protocol, which lays a promising method for producing metal nanoparticle arrays as templates for photonic devices, light harvesting for solar cells or mass fabrication of nanosensors based on SERS technology. Moreover, further deposition of metal cores with other morphologies, for example rods, stars or bimetallic core-shell Au@Ag[24] systems will extend significantly the range of potential applications based on such arrays.

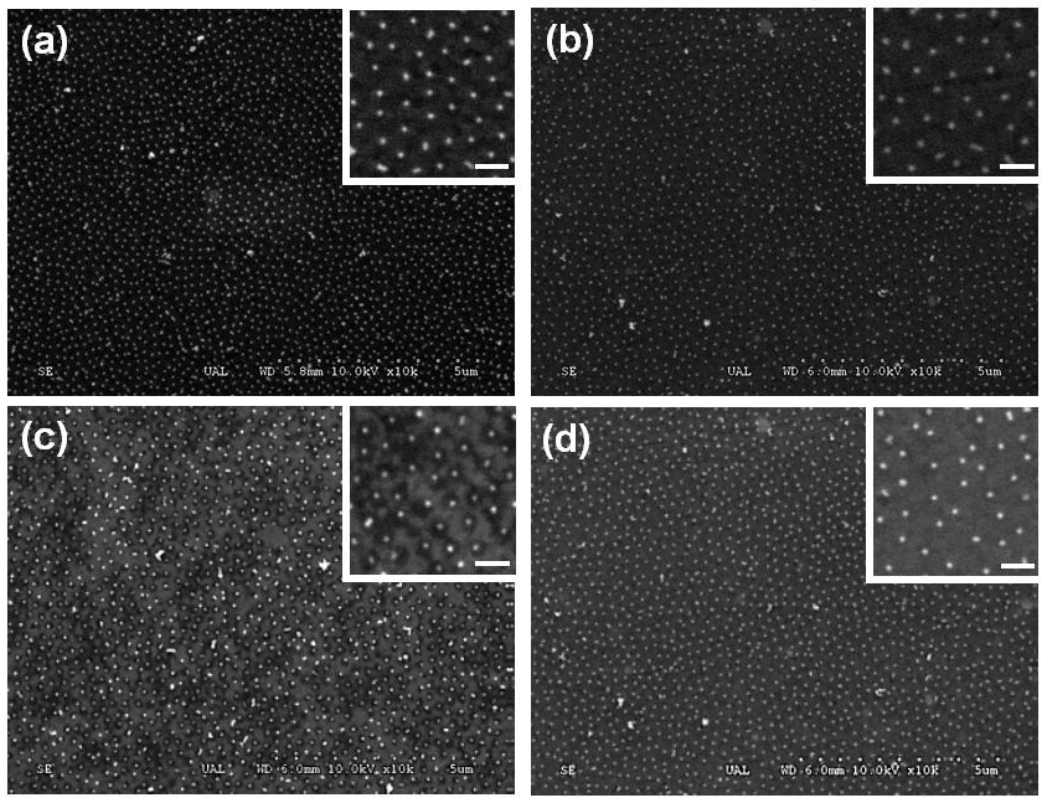

Fig. 8. SEM images of arrays deposited at *25ºC* after and exposure time of 0s (a), 20s (b), 40s (c) and 90s (d) to atmospheric plasma. Note the PNiPAM shell of the original Au@PNiPAM particles gradually degrading as the plasma exposure time is increased. For clarity, the insets are zoomed areas of each SEM image. The scale bar in the insets is 300nm.

### 4. Conclusions

Core-shell Au@PNiPAM nanoparticles are synthesized and nanoparticle arrays are successfully generated by drying bulk suspensions at different temperatures. PNiPAM shells are employed as spacers due to the polymer's thermal response and this is the key parameter responsible for determining the interparticle gap. Shell thickness varies by varying the drying temperature and sets the closest center-to-center distance between particles. These deposits arrange in ordered particle arrays which display structural properties similar to the ones of hexagonal lattices. Spatial order does not display a relevant length scale which is a hallmark of glassy colloidal systems and, in contrast, the angular configuration displays an increasing correlation length with increasing temperature, which is larger at larger temperatures. Despite the pair correlation function peaks are at distances multiple of the pair correlation primary peak, there is not a linear relation between peak positions at each temperature and the corresponding particle size in bulk, further emphasizing that the PNiPAM shell gradually decreases its elasticity by gradually increasing temperature and thus while gradually decreasing the PNiPAM-solvent affinity. This argument is also hinted by direct AFM imaging. All these features suggest that the shell is more elastic at lower temperatures and that the particle resembles more a hard-sphere colloid at higher ones. Au@PNiPAM particle deposits are exposed to atmospheric plasma and we successfully determine the conditions that allow a complete removal of the PNiPAM shell and therefore, leave the bare Au cores in arrays that lay off promising photonic applications, in particular towards improved efficiency of solar cells through enhanced plasmon resonance.

**Acknowledgments**

This work has been funded by the Spanish Ministerio de Economía y Competitividad/FEDER (project MAT2011-**28385**), Andalusian Government/FEDER (Project **P010-FQM 06104)** and EU-COST-Action CM1101.